# *n*-type electrical conduction in SnS thin films

Issei Suzuki[a,*], Sakiko Kawanishi[a,*], Sage R. Bauers[b], Andriy Zakutayev[b], Zexin Lin[a], Satoshi Tsukuda[a], Hiroyuki Shibata[a], Minseok Kim[c], Hiroshi Yanagi[c] and Takahisa Omata[a]

*a* Institute of Multidisciplinary Research for Advanced Materials, Tohoku University, 2-1-1 Katahira, Aoba-ku, Sendai, Miyagi 980-8577, Japan
*b* Material Science Center, National Renewable Energy Laboratory, Golden, CO 80401, U.S.A.
*c* Graduate Faculty of Interdisciplinary Research, University of Yamanashi, 4-4-37 Takeda, Kofu, Yamanashi 400-8510, Japan
*Issei Suzuki, issei.suzuki@tohoku.ac.jp
*Sakiko Kawanishi, s-kawa@tohoku.ac.jp

**Tin monosulfide (SnS) usually exhibits p-type conduction due to the low formation enthalpy of acceptor-type defects, and as a result, n-type SnS thin films have never been obtained. In this paper, we realize n-type conduction in SnS thin films by using radiofrequency-magnetron sputtering with Cl doping and a sulfur plasma source during deposition. Here, n-type SnS thin films are obtained at all the substrate temperatures employed in this paper (221–341 °C), exhibiting carrier concentrations and Hall mobilities of $2\times10^{18}$cm$^{-3}$ and 0.1–1cm$^2$V$^{-1}$s$^{-1}$, respectively. The films prepared without a sulfur plasma source, on the other hand, exhibit p-type conduction despite containing a comparable amount of Cl donors. This is likely due to a significant number of acceptor-type defects originating from sulfur deficiency in p-type films, which appears as a broad optical absorption within the band gap. We demonstrate n-type SnS thin films in this paper for the realization of SnS homojunction solar cells, which are expected to have a higher conversion efficiency than the conventional heterojunction SnS solar cells.**

## I. Introduction

Controlling the electrical properties of semiconductors by impurity doping is one of the most important techniques in semiconductor technologies. This technique could allow carrier-type conversion between the p- and n-type carriers and thus the formation of a pn homojunction. In nominally undoped semiconductors, the carrier type is generally determined by the defects with a low formation enthalpy and low ionization energy, and this often makes carrier-type conversion difficult. The history of GaN-based blue light-emitting diodes (LEDs) is a good example. Carrier-type conversion of intrinsically n-type GaN to p-type by impurity doping had been the primary obstacle, and therefore, its achievement was the biggest breakthrough for the realization of blue LEDs [1], leading to a Noble Prize in physics in 2014.

Tin monosulfide (SnS) is a promising material for the solar light absorber for thin film solar cells because it is composed of abundant and nontoxic elements and exhibits an appropriate optical property for solar cells. SnS usually exhibits p-type conduction due to the intrinsic defects. Under S-rich conditions, the acceptor-type Sn vacancy ($V_{Sn}$) is easily formed. Under S-poor conditions, Sn tends to migrate into the vacant S sites, and then acceptor-type VSn and Sn antisites ($Sn_S$) are easily formed. Regardless of the conditions, S vacancy ($V_S$) is an ambipolar defect with deep states that can trap both electrons and holes [2–4]. Therefore, there has been a long challenge to realize the n-type SnS thin film by impurity doping. However, substitutional cation doping of $Sn^{2+}$ with trivalent cations, such as $Sb^{3+}$ and $Bi^{3+}$, did not lead to n-type conduction [4–6]. The fabrication of n-type SnS thin films thus remained elusive to date.

Recently, in SnS bulk pellets and single crystals, it was found that substitutional anion doping of $S^{2-}$ with halide ions, such as $Cl^-$ and $Br^-$, successfully achieved n-type electrical conduction [7–11]. This doping approach is more suitable for practical use than the solid solutions between SnS and PbS ($Sn_{1-x}Pb_xS$), which are the only SnS-containing n-type semiconductors reported before the halogen-doped SnS [12,13] and unfortunately contains a large amount of toxic Pb ($x$>0.2).

Compared with bulk n-type SnS, the difficulty in fabricating n-type SnS thin films by physical vapor deposition (such as sputtering, pulsed laser deposition, or thermal evaporation) is likely related to the preferential formation of intrinsic defects under the deposition atmosphere. Because the vapor pressure of elemental sulfur is 11 orders of magnitude higher than that of elemental tin at 500 °C [14], SnS thin films prepared in the open system of physical vapor deposition are probably more prone to sulfur deficiency than the bulk SnS prepared in a substantially closed system. Accordingly, it is hypothesized that acceptor-type defects ($V_{Sn}$ and $Sn_S$) and ambipolar defects ($V_S$), both caused by sulfur deficiencies, compensate the electrons introduced by halogen doping, preventing the halogen-doped SnS thin films from exhibiting n-type conduction. Innovative thin film deposition methods are needed to address this challenge.

Herein, n-type SnS thin films were fabricated by radiofrequency (RF) magnetron sputtering together with an RF sulfur plasma source. The additional sulfur plasma supplied to the films during deposition increased the sulfur content of the resulting films. It was found that using the sulfur plasma source during deposition significantly suppressed the formation of acceptor-type defects in the SnS films, presumably $V_{Sn}$ and $Sn_S$, as well as ambipolar $V_S$. This is the key to realize n-type conduction in Cl-doped SnS thin films. The fabrication technique demonstrated here for n-type SnS thin films is expected to revolutionize SnS solar cells, from heterojunctions developed over the last two decades to homojunctions with much higher efficiency.

## II. Experimental

SnS thin films were fabricated by RF-magnetron sputtering. To investigate the substrate temperature dependence of the physical properties of SnS thin films, a $SiO_2$ glass substrate (50 × 50 mm) with a temperature gradient across it (221–341 °C) [15] was used (for the spatial distribution of substrate temperature, see Fig. S1 in the Supplemental Material [16]). The deposition time was 30 min. A 2-inch-diameter piece of undoped SnS (99.99%, Advanced Engineering Materials) was used as the sputtering target. Before thin film deposition, Cl was supplied to the deposition vacuum chamber by exposing a Cl-containing SnS target [Cl/(SnS + Cl) = 0.1, 99.99%, Advanced Engineering Materials] to Ar plasma for 30 min under the same conditions as the thin film deposition, during which a temporary substrate was set on the substrate holder. Then the deposition chamber was evacuated to its base pressure ($<5\times10^{-4}$ Pa), and the substrate was replaced. A sulfur plasma source (RF solid plasma source RFK30, Oxford Applied Research) was used in conjunction with RF sputtering to avoid sulfur deficiency in the resulting thin films. Further details of the thin film fabrication (such as the specific sputtering conditions, configuration of the deposition equipment, and operating conditions of the sulfur plasma source) are available in Sec. S2 in the Supplemental Material [16] together with details of the thin film characterization methods.

## III. Results and discussion

The thin film prepared at a substrate temperature ($T_{sub}$) of 333 °C is firstly discussed as a representative sample. As shown in Fig. 1(a), all x-ray diffraction (XRD) peaks of the obtained films are indexed as α-SnS (see Fig. S3 in the Supplemental Material [16] for the crystal structure), and no impurity phases such as polymorphs of SnS, $SnS_2$, and $Sn_2S_3$ were observed whether sulfur plasma was supplied or not during deposition. The intense peak observed at around 2θ = 32 ° corresponds to the 400 diffraction peak of α-SnS. This peak has a much stronger intensity than the others, indicating that the obtained SnS films exhibited preferential (100) orientation at $T_{sub}$=333 °C. This result is consistent with the orientation mappings obtained by electron backscattered diffraction (EBSD, see Fig. S4 in the Supplemental Material [16]).

The chemical compositions of the films deposited with and without sulfur plasma supply were evaluated by x-ray fluorescence spectroscopy (XRF) [17]. Their compositions were close to the stoichiometry of SnS [S/(Sn + S) = 0.5], as indicated in Table I. Figure 1(b) shows the depth profile of Cl concentration and the intensities of secondary ions of $Sn^-$, 34 $S^-$, and $Si^-$ determined by time-of-flight secondary ion mass spectroscopy (TOF-SIMS). The concentrations of Sn, S, and Cl were almost uniform in the thickness direction, regardless of sulfur plasma supply. The Cl concentrations shown in Table I were obtained by averaging the bulk part of the n- and p-type SnS thin films [70–110 nm from the surface in Fig. 1(b)]. The in-plane analysis by TOF-SIMS also showed uniform distribution of each element over a film area of 100 × 100 μm, thereby excluding the possibility of segregation due to impurity phases (see Fig. S5 in the Supplemental Material [16]). All results of phase identification and chemical compositional analysis support that the obtained films consist of a uniform single-phase α-SnS.

In Fig. 1(c), both surface scanning electron microscopy (SEM) and atomic force microscopy (AFM) observations indicate that the obtained SnS films are composed of almost circular grains with a diameter of 10 to 200 nm regardless of sulfur plasma supply, while grain size tends to be slightly smaller in the film deposited with sulfur plasma supply (<100 nm). The obtained SnS thin films consisted of much finer grains as SnS thin films prepared by RF-magnetron sputtering than the previous reports [18,19].

The electrical properties are also summarized in Table I. The SnS thin film prepared with sulfur plasma supply showed negative values of both Hall and Seebeck coefficients, which indicate n-type conduction. On the other hand, the film prepared without sulfur plasma supply had positive values for both coefficients, indicating p-type conduction. Furthermore, the Fermi level shift in the electronic structure was directly observed by bulk-sensitive x-ray photoelectron spectroscopy (XPS) measured with monochromatic Ag Lα excitation (hν = ~ 3 keV). As shown in Fig. 2(a), the binding energies of the valence band edge, S 2p, and Sn 3d peaks of the film deposited with sulfur plasma supply (the n-type film) shifted downward by ~0.3 eV compared with that without sulfur plasma (p-type), which is direct evidence that the Fermi level of the n-type film is 0.3 eV closer to the conduction band edge than the p-type film. These results show that the presence or absence of sulfur plasma supply determines the carrier type of the thin films.

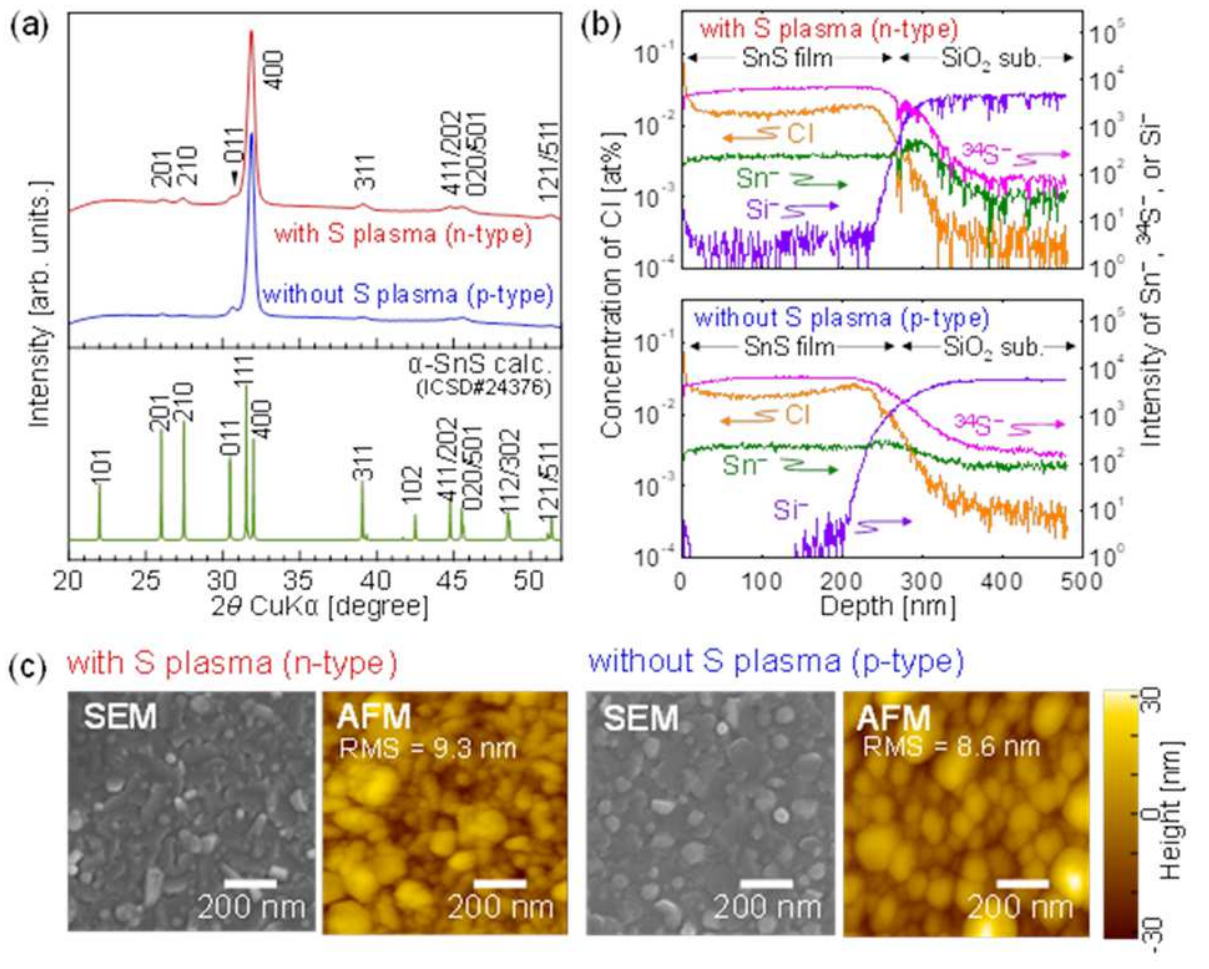

**Figure 1**. Structural, compositional, and morphological analysis of Cl-doped SnS thin films deposited with and without a sulfur plasma supply. (a) X-ray diffraction (XRD) profiles of films deposited at $T_{sub}$=333 °C, together with a reference pattern of α-SnS (ICSD#24376). (b) Depth profile of Cl concentration and normalized intensities of secondary ions of $Sn^-$, $34S^-$, and $Si^-$ in the films deposited at Tsub=333 °C. (c) Scanning electron microscopy (SEM) and atomic force microscopy (AFM) images of the surface of SnS thin films deposited at $T_{sub}$=333 °C. The root-mean-square (RMS) roughness is also shown. Additional SEM and AFM images are available in Fig. S6 in the Supplemental Material [16].

**Table 1**. Physical properties of SnS thin films prepared at $T_{sub}$ = 333 °C with and without sulfur plasma supply. All results were obtained at room temperature, except that the Seebeck coefficients were measured at 25–50 °C.

| | SnS thin films deposited | |
|---|---|---|
| | with sulfur plasma | without sulfur plasma |
| Thickness [nm] | 250 | 190 |
| Atomic ratio, S/(Sn+S) | 0.507 | 0.498 |
| Cl concentration [cm$^{-3}$] | $3.7\times10^{18}$ | $4.6\times10^{18}$ |
| Hall coefficient [cm$^3$ C$^{-1}$] | $-2.6\times10^{0}$ | $+4.8\times10^{3}$ |
| Seebeck coefficient [μV K$^{-1}$] | $-6.3\times10^{1}$ | $+1.4\times10^{4}$ |
| Carrier type | n-type | p-type |
| Electrical conductivity [S cm$^{-1}$] | $9.2\times10^{-2}$ | $3.1\times10^{-4}$ |
| Carrier concentration [cm$^{-3}$] | $2.4\times10^{18}$ | $1.4\times10^{15}$ |
| Hall mobility [cm$^2$ V$^{-1}$ s$^{-1}$] | $2.4\times10^{-1}$ | $1.5\times10^{0}$ |

It is interesting to note that n-type conduction in SnS thin films could be achieved by supplying sulfur plasma during deposition together with Cl doping but not by Cl doping alone. The carrier concentration of the n-type film is close to the Cl donor concentration (Table I), suggesting that carrier electrons are generated by Cl doping, and the activation efficiency of the Cl donors in this film is as high as ~65%. The film prepared without sulfur plasma had a similar Cl concentration; however, it did not exhibit n-type conduction. This clearly indicates that n-type conduction in SnS thin films cannot be achieved only by incorporating Cl donors, likely due to compensation by intrinsic defects.

Other important features and clues related to Cl incorporation and defect compensation were observed in the optical absorption spectra of these films. In Fig. 2(b), in addition to the steep slope starting from 1.3 eV corresponding to the fundamental absorption of SnS, a broad absorption centered at ~1.1 eV was much more obvious in the film deposited without supplying sulfur plasma (p-type film) than the film deposited with it (n-type film). This broad absorption is distinguished from interband transitions in the $h\nu$ vs $(\alpha h\nu)1/r$ plots, where $h\nu$, $\alpha$, and $r$ are the photon energy, absorption coefficient, and an index related to the interband transition type, respectively [20] (see Fig. S8 in the Supplemental Material [16]). This indicates the presence of many in-gap states in the p-type films deposited without sulfur plasma supply. Since this absorption was not obvious in the n-type film prepared with sulfur plasma supply, it should be related to the defects associated with the sulfur deficiency.

According to first-principles calculations [2–4], $Sn_S$ and $V_{Sn}$

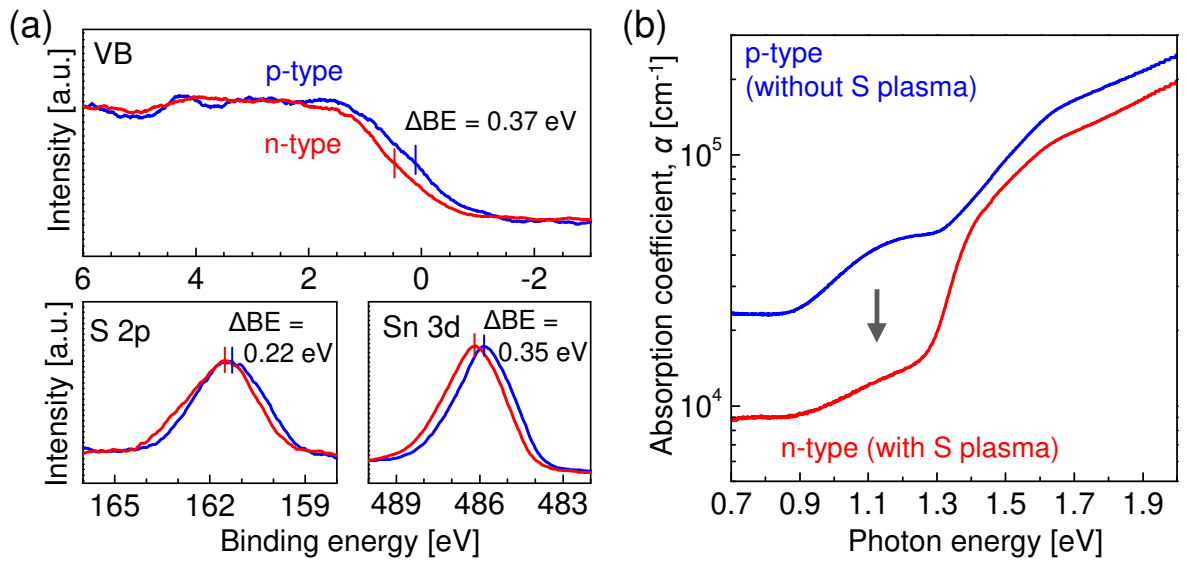

**Figure 2.** (a) X-ray photoelectron spectroscopy (XPS) spectra (excitation source: Ag Lα) of SnS films deposited at $T_{sub}$=222 °C for the valence band (VB), S 2p, and Sn 3d core levels. Differences in the binding energies between n- and p-type films (ΔBE) are shown in each plot. A total resolution of ~0.5 eV for the measurement resulted in significant spectral broadening. The displayed spectra were obtained by the moving average of the original results, which are available in Fig. S7 in the Supplemental Material [16]. Other characteristics of this film ( $T_{sub}$=222 °C) are discussed in the last half of main text. (b) Absorption spectra of the n- and p-type SnS thin films prepared at $T_{sub}$=333 °C. The broad peak centered at 1.1 eV is significant only for the p-type film.

are the most likely defects formed under sulfur-deficient conditions, and they act as acceptors. In addition, $V_S$ is an ambipolar defect that can also compensate donors when the Fermi level is high in the gap. Therefore, the broad and strong absorption in the p-type film ought to be related to the in-gap states due to $Sn_S$, $V_{Sn}$, $V_S$, or their complex with Cl impurities (such as $Sn_S$-$Cl_S$, $V_{Sn}$-$Cl_S$, or $V_S$-$Cl_S$). The SnS thin film will not exhibit n-type conduction even with halogen doping if such defects are present at a high concentration. Sulfur deficiency is commonly observed in SnS thin films prepared by physical vapor deposition [21–23]. Postanneal of the sulfur-deficient SnS thin films in an atmosphere with high sulfur activity, such as H2S, cannot effectively reduce the concentration of holes (i.e., acceptor-type defects) [24]. Therefore, it is supposed that supplying highly reactive sulfur (i.e., sulfur plasma) during film deposition, which was employed in this paper, is essential for realizing n-type conduction in Cl-doped SnS thin films.

Figure 3(a) plots the temperature-dependent carrier concentration in the n- and p-type films. Carrier concentration of the p-type film showed thermally activated behavior with an activation energy of $E_a$ = 230 meV, which is comparable with those of undoped p-type SnS thin films from previous reports ($E_a$=~300 meV [25,26]). The n-type film exhibited almost no temperature dependence, indicating that the film is an almost degenerated semiconductor. Like the control of carrier concentration in n-type SnS bulk ceramics by the Cl doping level, [11] the carrier concentration of n-type films should also be controllable by the amount of incorporated Cl donor. Figure 3(b) shows temperature dependence of the Hall mobility for the films. The Hall mobility decreases with decreasing temperature for both n- and p-type SnS thin films. The $\log(\mu_{Hall}T^{1/2})$–$T^{-1}$ plot exhibits a straight line, suggesting that grain boundary scattering dominates carrier transport in the films [27]. This is consistent with the fact that carrier transport is not dominated by ionized scattering or phonon scattering in these films (see

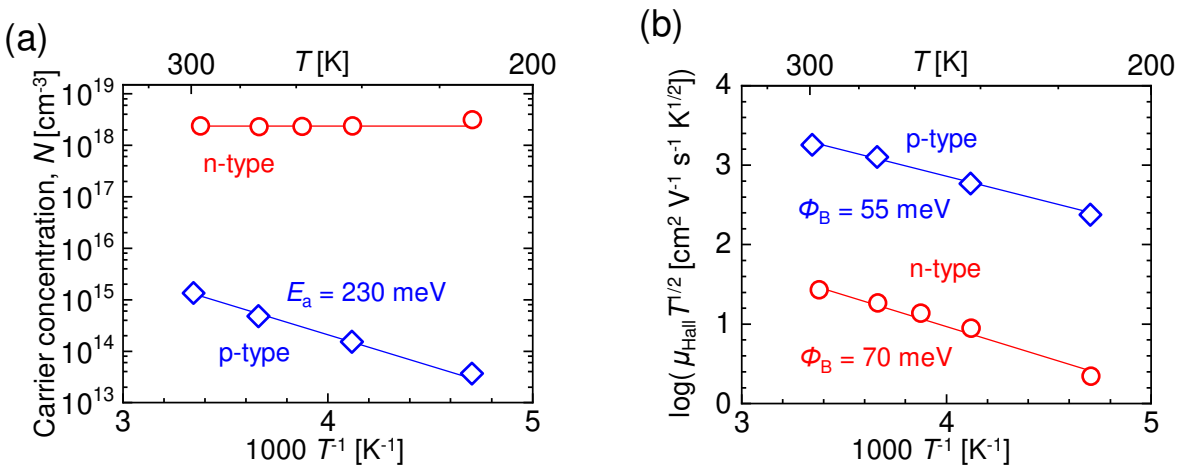

**Figure 3.** a) Temperature dependence of carrier concentration and b) $\mu_{Hall}T^{1/2}$ of SnS thin films deposited at $T_{sub}$ = 333 °C. The activation energy ($E_a$) and grain boundary potential barriers ($\Phi_B$) were estimated according to the following formulas: $N = N_0\exp(-E_a/kT)$, where $N_0$ is a prefactor and $k$ is Boltzmann constant; and $\mu_{Hall} = Le(1/2\pi m^*kT)^{1/2}\exp(-\Phi_B/kT)$, where $L$ is the grain size, $e$ is elementary charge, and $m^*$ is the effective mass. The solid lines were added to guide the eye.

Fig. S9 in the Supplemental Material [16] for double-log plots of Hall mobility and temperature). The calculated potential barriers at the grain boundary ($\Phi_B$) are 70 and 55 meV for n- and p-type SnS thin films, respectively, which are of the same order of magnitude as values previously reported for undoped p-type SnS thin films that are dominated by grain boundary scattering ($\Phi_B$=50–70 meV [28,29]). The main reason for the dominance of grain boundary scattering in the obtained films is their small grain sizes [which were discussed above, see Fig. 1(c)], which lead to many grain boundaries. It has been reported in several SnS studies that the carrier mobility of SnS thin films increases with increasing grain size [18,30,31], which is also a general property of polycrystalline thin films with dominant grain boundary scattering [32]. The morphology of the SnS thin films strongly depends on the deposition conditions, especially the sputtering pressure. The films deposited at relatively low pressures ≤~1 Pa, as the pressure used in this paper (0.35 Pa), tend to be flatter with finer grains, while those deposited at high pressures (≥~1 Pa) possess larger grains and rougher surfaces [18,19]. Therefore, optimizing the sputtering conditions would allow the deposition of films with larger grains and correspondingly higher carrier mobility.

Substrate temperature dependence (221–341 °C) of physical properties of SnS thin films was also investigated. The films prepared with and without sulfur supply exhibited n- and p-type conduction, respectively, regardless of the substrate temperature (for the electrical properties of each sample, see Fig. S10 and Tables S4 and S5 in the Supplemental Material [16]). The obtained films were of single-phase α-SnS and exhibited preferential (100) orientations regardless of the substrate temperature (see Figs. S11 and S12 in the Supplemental Material [16] for XRD profiles of out-of-plane (2θ) and in-plane ($2\theta_\chi$), respectively). The primary peaks corresponding to 400 diffractions shifted toward lower diffraction angle with decreasing the substrate temperature, indicating that the lattice constant in the $a$ axis, $a_0$, increased with decreasing the substrate temperature, as shown in Fig. 4(a). The thicknesses of the obtained films were 250–260 and 190–200 nm when deposited with and without sulfur plasma, respectively (see details in Fig. S13 in the Supplemental Material [16]), and therefore, supplying sulfur plasma increases the deposition rate by ~30%.

Figure 4(b) plots the chemical composition of SnS thin films against the substrate temperature. The composition is almost independent of the substrate temperature, and supplying sulfur plasma slightly increased sulfur content in the resulting films by ~1%. The broad optical absorption related to the defect levels associated with acceptor-type defects [which were discussed above, see Fig. 2(b)] was observed in all p-type SnS films at similar intensities, whereas they were hardly visible in the n-type films at any substrate temperature (see Fig. S8 in the Supplemental Material [16]). Therefore, changing the substrate temperature cannot suppress the formation of acceptor-type defects due to sulfur deficiency, while supplying sulfur plasma is always effective in suppressing these defects and leads to n-type conduction at any substrate temperature employed in this paper.

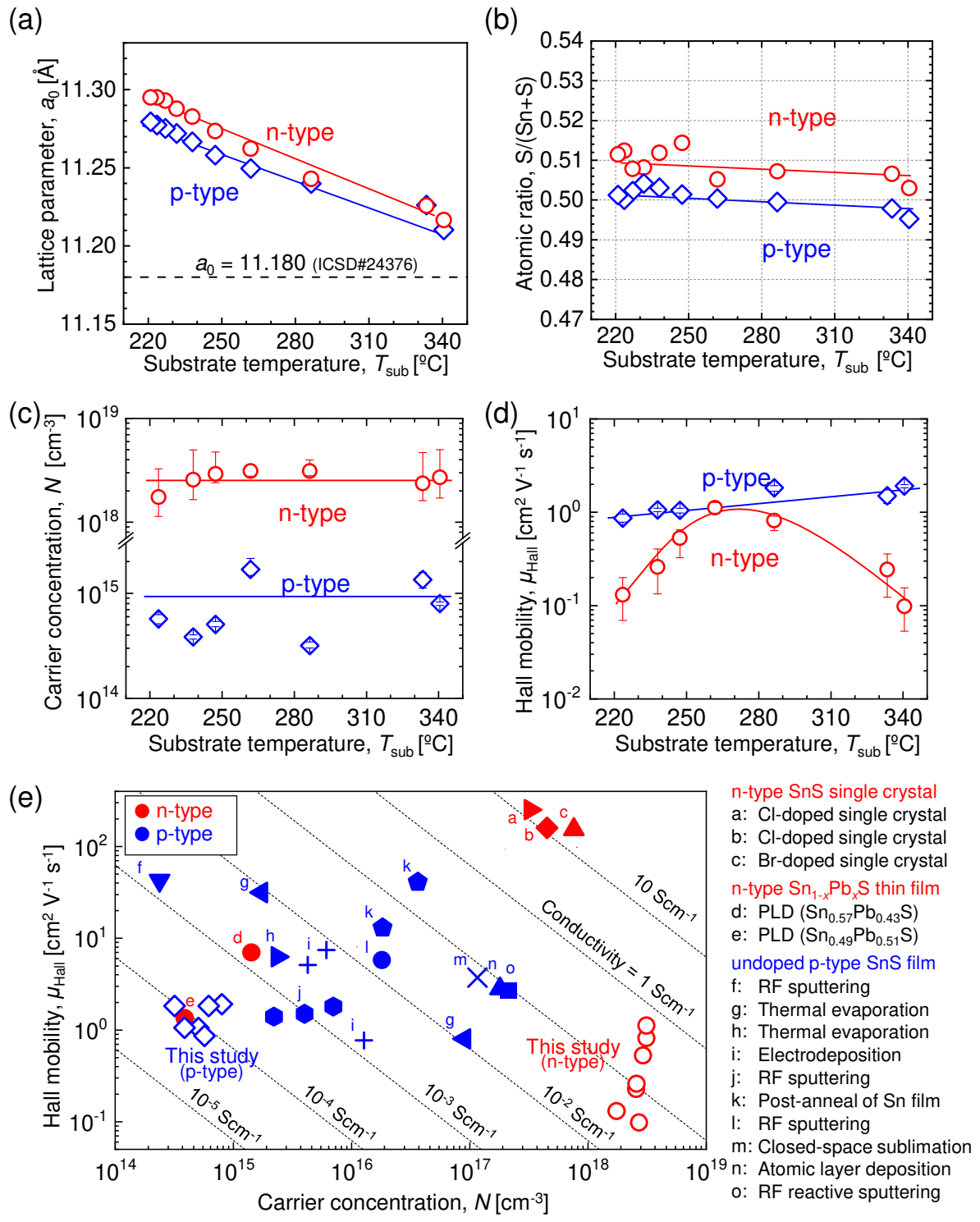

**Figure 4.** Substrate temperature dependence of the various properties of the obtained SnS thin films. (a) Lattice parameter of a0 of the obtained thin films determined by 400 diffraction peaks (for the x-ray diffraction (XRD) profiles, see Fig. S11 in the Supplemental Material [16]). (b) Chemical composition determined by x-ray fluorescence spectroscopy (XRF). (c) Carrier concentration and (d) Hall mobility of SnS thin films measured at room temperature. All solid lines in the figures were added to guide the eye. Detailed data on the electrical properties are available in Fig. S10 and Tables S4 and S5 in the Supplemental Material [16]. (e) Comparison of electrical properties of n- and p-type SnS thin films obtained in this paper and those of previously reported n- and p-type SnS, and n-type $Sn_{1-x}Pb_xS$ solid solutions. The literature results came from aRef. [8], b,cRef. [9], d,eRef. [13], fRef. [19], gRef. [42], hRef. [43], iRef. [44], jRef. [18], kRef. [45], lRef. [41], mRef. [21], nRef. [46], and oRef. [47].

Figures 4(c) and 4(d) show the substrate temperature dependence of the electrical properties for both n- and p-type SnS thin films. The carrier concentrations of n-type films did not depend on the substrate temperature (~$2\times10^{18}$ $cm^{-3}$), implying that all these films had comparable Cl concentration. The Hall mobility of the p-type thin films did not depend on the temperature either, whereas the n-type films exhibited a convex trend with a maximum at $T_{sub}$~260 °C. Their carrier type, chemical composition, and number of acceptor-type defects were not affected significantly by the substrate temperature but rather by whether sulfur plasma was supplied during the deposition. These results indicate that the fabrication process of n-type SnS thin films demonstrated here could be applied under a wide range of sputtering conditions. The n-type SnS thin films obtained in this paper have a high carrier concentration and low Hall mobility compared with the previously reported p-type films and n-type bulk single crystals, as shown in Fig. 4(e). By tuning the deposition conditions, the proposed fabrication process using sulfur plasma supply has high potential for realizing n-type SnS thin films with characteristics that are favorable for device applications, for instance, lower carrier concentration and longer minority carrier diffusion length.

Thin film SnS is expected to be a solar light absorber for next-generation solar cells because of its abundant and nontoxic constitutional elements and suitable optical properties for photovoltaic applications. Due to a lack of n-type SnS thin films, SnS thin film solar cells based on the heterojunction structure between p-type SnS and other n-type semiconductors, such as CdS [33–35] or Zn(O,S) [24,36,37], have been intensively studied in the last two decades. However, due to their unfavorable conduction band offset [35,38] and possible Fermi level pinning induced by defects at the heterojunction interface [34], SnS heterojunction solar cells have a relatively low open-circuit voltage ($V_{OC}$ = 300–400 mV) compared with its band gap energy (~1.1 eV), leading to low conversion efficiencies ($\eta$) of only 4–5% at the most [39]. The pn homojunction of SnS employing n-type SnS thin film developed in this paper is ideally free from these problems, making it the most promising way to improve the open-circuit voltage and conversion efficiency [13,40]. This expectation is supported by the recent study of a prototype SnS homojunction solar cell with halogen-doped n-type SnS single crystal and p-type SnS thin film exhibited $V_{OC}$ comparable with the highest values for heterojunctions, without optimizing the device structure or fabrication process [41].

In addition, supplying sulfur plasma during deposition is beneficial for the fabrication of high-quality undoped p-type SnS thin films because it can effectively suppress the formation of acceptor-type defects. Therefore, undoped p-type SnS thin films fabricated under a Cl-free atmosphere would have a lower concentration of defects that act as effective recombination centers and be suitable for solar cell applications. In other words, a sulfur plasma supply can be employed to produce high-quality thin films with both n- and p-type conduction. Consequently, this method will surely be key for realizing highly efficient SnS homojunction solar cells.

## IV. Conclusion

In this paper, we realized n-type electrical conduction in SnS thin films by supplying sulfur plasma during the deposition together with Cl doping. Without sulfur plasma supply, the prepared SnS thin films exhibited p-type conduction despite containing a comparable amount of Cl donors, and the broad optical absorption of those films would be attributed to the acceptor-type defects originating from sulfur deficiency. Accordingly, both Cl doping and the supply of highly reactive sulfur plasma during deposition are essential to realize n-type conduction in SnS thin films. The reported fabrication method can be applied to a wide range of sputtering conditions since n-type conduction was obtained regardless of the substrate temperature. So far, the conversion efficiency of SnS heterojunction solar cells is stagnant at low values (4–5%). Methods to fabricate n-type SnS thin films will enable SnS pn homojunctions and contribute to a change in the SnS solar cell technology from heterojunctions to homojunctions to realize high conversion efficiency.

### Acknowledgement

The authors thank Profs. Chichibu and Shima at Tohoku University for their help with the Hall measurement equipment. The authors are deeply grateful to Ms. Shishido and Mr. Magara at Tohoku University for their professional technical support in the TOF-SIMS analysis. This paper resulted from the collaboration between Tohoku University, the University of Yamanashi, and National Renewable Energy Laboratory (NREL) under the support of Fostering Joint International Research (B) (Grant No. 18KK0133). The chalcogenide synthesis facility at NREL is funded by the Liquid Sunlight Alliance, which is supported by the U.S. Department of Energy (DOE), Office of Science, Office of Basic Energy Sciences, Fuels from Sunlight Hub under Award No. DE-SC0021266. NREL is operated by Alliance for Sustainable Energy, LLC, for the DOE under Contract No. DE-AC36-08GO28308. This paper was also partly supported by a Grant-in-Aid for Scientific Research (B) (Grant No. 19H02430), the Murata Science Foundation, IMRAM project funding, and the Research Program of "Five-star Alliance" in "NJRC Mater. & Dev." The views expressed in the article do not necessarily represent the views of the DOE or the U.S. Government.

### Supplemental Material

Detailed information on thin-film fabrication and characterization, temperature gradient in the substrate, crystal structure, EBSD mapping, in-plane analysis by TOF-SIMS measurement, additional SEM and AFM images of thin film morphology, XPS with Ag-L excitation, optical properties and determination of band gaps, double log plots of Hall mobility vs. temperature, detailed dependence of electrical properties on substrate temperature, XRD profiles of SnS thin films deposited at various substrate temperatures and thickness of thin films.

Supplemental Material

**N-Type Electrical Conduction in SnS Thin films**

*Issei Suzuki*, Sakiko Kawanishi*, Sage R. Bauers, Andriy Zakutayev, Zexin Lin, Satoshi Tsukuda, Hiroyuki Shibata, Minseok Kim, Hiroshi Yanagi, and Takahisa Omata*

## Section S1. Temperature gradient in the substrate

➢ **Temperature gradient**

The dependence of film properties on the substrate temperature was studied using a single large sample prepared with a temperature gradient (221–341 °C) within the substrate, as shown in **Figure S1**.

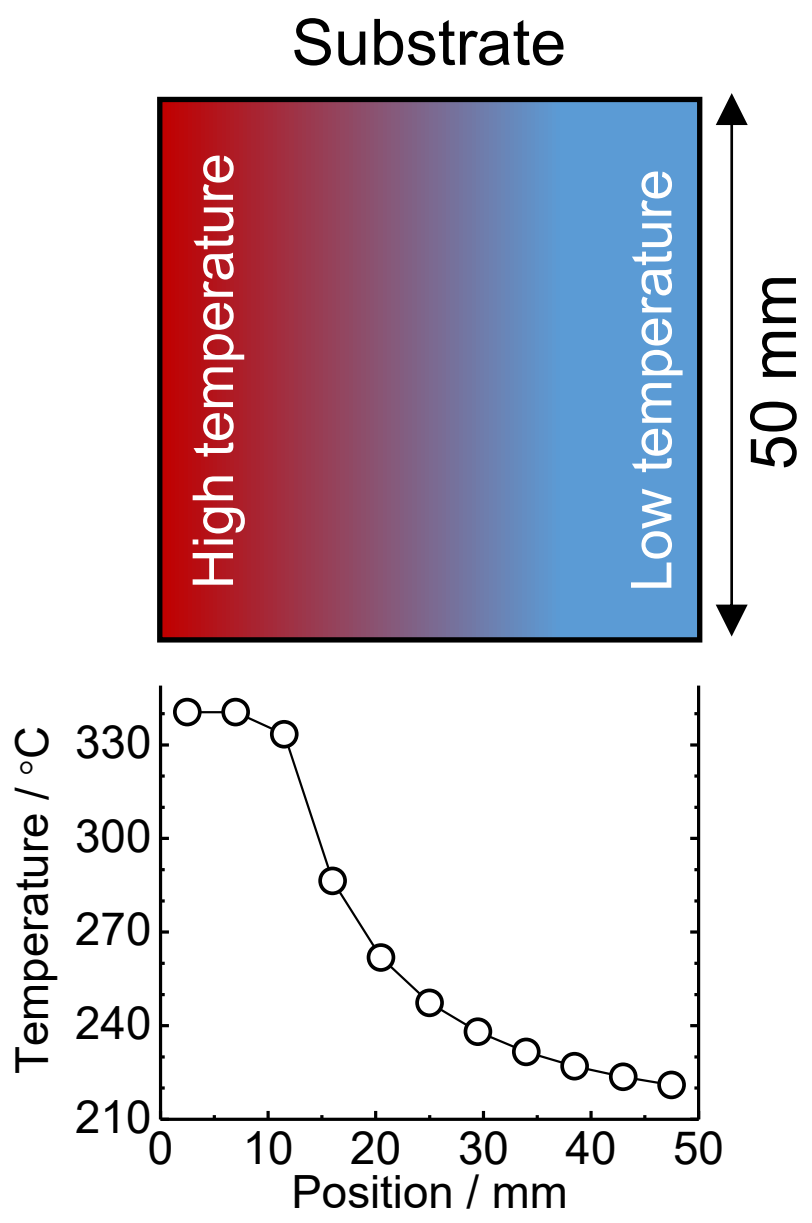

**Figure S1.** Schematic diagram of temperature gradient in the substrate and the actual temperatures.

## Section S2. Detailed information on thin-film fabrication and characterization

### ➢ Detailed sputtering conditions

**Table S1.** Sputtering conditions.

| | |
|---|---|
| Base pressure / Pa | < $5\times10^{-4}$ |
| Gas flow | Ar, 10 sccm |
| RF power / W | 40 |
| Deposition time / min | 30 |
| Substrate temperature / °C | Gradient (Figure S1) |
| Sputtering pressure / Pa | 0.35 |

### ➢ Structure of sulfur plasma source and its operating conditions

The sulfur plasma source was installed next to the sputtering cathode in the deposition chamber, as shown in **Figure S2(a)**. It feeds sulfur plasma to the deposition area during deposition. Sulfur is heated and evaporated at the lower part of the sulfur plasma source, as shown in **Figure S2(b)**. The evaporated sulfur then conveyed to the RF coil. The operating conditions are summarized in **Table S2**.

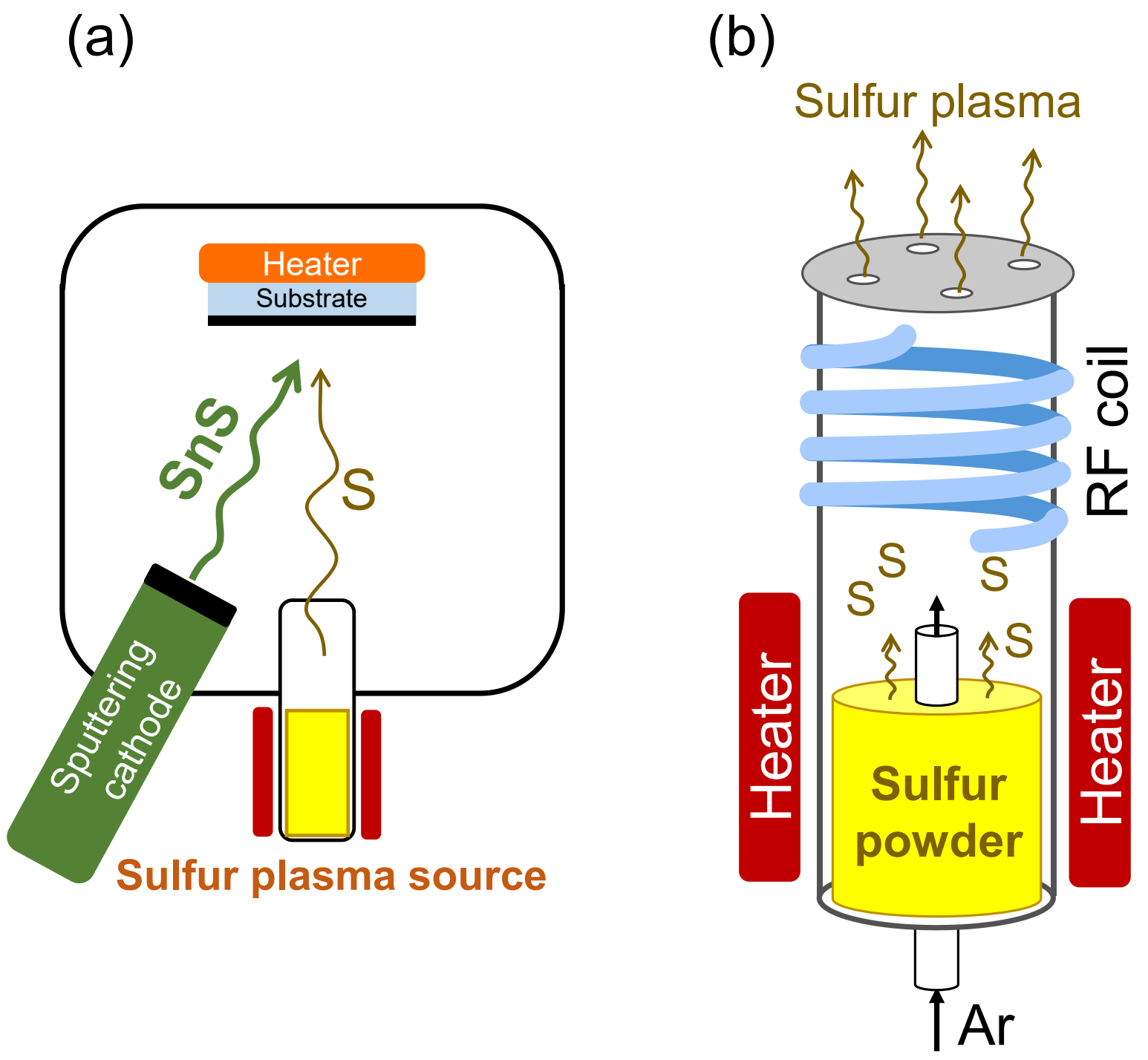

**Figure S2.** Schematics of (a) the sulfur plasma source installed in the sputtering chamber and (b) structure of the sulfur plasma source.

**Table S2.** Operating conditions of the sulfur plasma source.

| | |
|---|---|
| Heater for sulfur source | 75 °C |
| RF power | 50 W |
| Gas flow | Ar, 6 sccm |

## Characterization

- **XRD**

The samples were mapped at 11 points to observe the crystal phases as a function of substrate temperature. The phases in the films were identified by X-ray diffraction (XRD) with Cu Kα radiation (Bruker, D8 Discover, MA, US).

- **Thickness**

Thickness of the films were determined by X-ray fluorescence spectroscopy (XRF, Matrix Metrologies MaXXi 5, CA, US).

- **TOF-SIMS**

The depth profile and the in-plane distribution of elements of the films were evaluated by time-of-flight secondary ion mass spectroscopy (TOF-SIMS, TOF.SIMS 5, IONTOF GmbH, Germany). To determine the Cl concentration from the measured ion currents, an n-type Cl-doped SnS single crystal with a Cl concentration of ~0.40at% was used as a reference.[S3]

- **Hall coefficient**

Electrical properties of the films were evaluated by the Hall coefficient and electrical conductivity. The temperature dependence of these parameters in the range of 210–300 K was measured by the van der Pauw method (ResiTest8300, Toyo Corporation, Japan). Thin films of Au (thickness: approximately 100 nm) were deposited as electrodes by desktop magnetron sputtering (JFC-1600, JEOL, Japan) at 20 mA. In order to reduce the effect of heating from sputtering and avoid the associated chemical reactions, a total of 16 cycles were repeated, each consisting of 10 s Au sputtering followed by 20 s of pause for cooling down.

- **Seebeck coefficient**

The Seebeck coefficients were measured using a homemade device. One end of the thin film was placed in contact with a ceramic heater, and the electromotive force was measured by a multimeter at 25–45 °C with a temperature difference of 0–20 °C between both ends of the film sample. Au thin films were used as electrodes.

➢ **XPS with Ag-Lα excitation source**

XPS spectra of the films were recorded using a photoemission spectrometer with a hemispherical electron analyzer (Axis Ultra DLD, Kratos Analytical, U.K.) at room temperature and excited by monochromatic Ag Lα radiation (hν = 2984.2 eV). The total resolution, which was evaluated from the Fermi edge of the Au films, was ~0.5 eV. The Fermi level of the films was calibrated by shifting the C 1s peak using the surface contamination peak at 285.0 eV. Since the measurement was not in-situ, that is, the samples were exposed to the air before measurement, the spectra were affected to a certain extent by the top contamination and oxidation layers.

➢ **Transmittance spectra**

Transmittance spectra in the wavelength range from 400 to 2000 nm were recorded using a UV-Vis-IR spectrophotometer (U-4100, Hitachi High-Tech, Japan). Interference fringes in the transmittance spectra due to reflections were removed using Fresnel's equation.

➢ **SEM and AFM**

Surface morphological observations of the thin films were carried out using field-emission scanning electron microscopy (FE-SEM, JSM-7800F, JEOL Ltd.) and atomic force microscopy (AFM, SPI-4000, SEIKO Instruments Inc., Japan). The FE-SEM observations were performed using a gentle beam mode at an acceleration voltage of 3 or 5 kV under a sample bias of 2 kV.

## Section S3. Crystal structure and axis setting

**Figure S3** schematically illustrates the crystal structure of α-SnS, which is also called GeS-type SnS. The setting of the SnS axis varies depending on the literature. In this study, the stacking direction of the SnS layers is set to the a-axis.

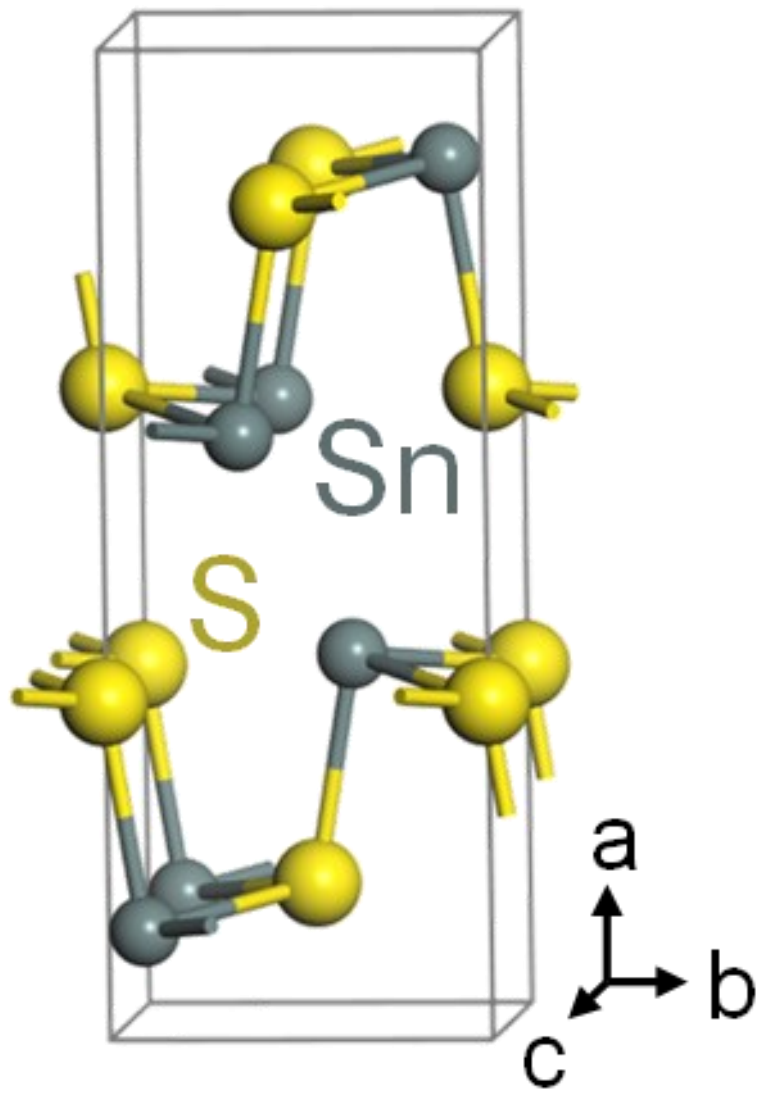

**Figure S3.** Crystal structure of α-type SnS.

## Section S4. EBSD mapping

The crystal orientations of films deposited at a substrate temperature ($T_{sub}$) of 333 °C were analyzed by electron back-scatter diffraction (EBSD) (Nordlys Nano, Oxford Instruments, UK) at an accelerating voltage of 10 kV. The obtained Kikuchi patterns were analyzed with the reference crystal structure of α-SnS (ICSD# 24376). The upper panels of **Figure S4(a,b)** show the inverse pole figure (IPF) maps for the normal direction (ND) and rolling direction (RD), and the lower panels are pole figures of {100} and {010} for the areas in the maps. It should be noted that no impurity phases (such as $SnS_2$, $Sn_2S_3$, and other polymorphs of SnS) were detected anywhere for both n-type and p-type thin films, and the black-colored area in the IPF maps did not match with any patterns due to the weak band contrast. In the ND maps, most of the grains are blue-colored for both n-type and p-type films, corresponding to the (100)-orientation. The small distributions around the center of the {100} pole figures indicate that the grains are slightly inclined with each other. The RD maps and {010} pole figures confirm that the two types of thin film showed no preferred orientation in the in-plane direction.

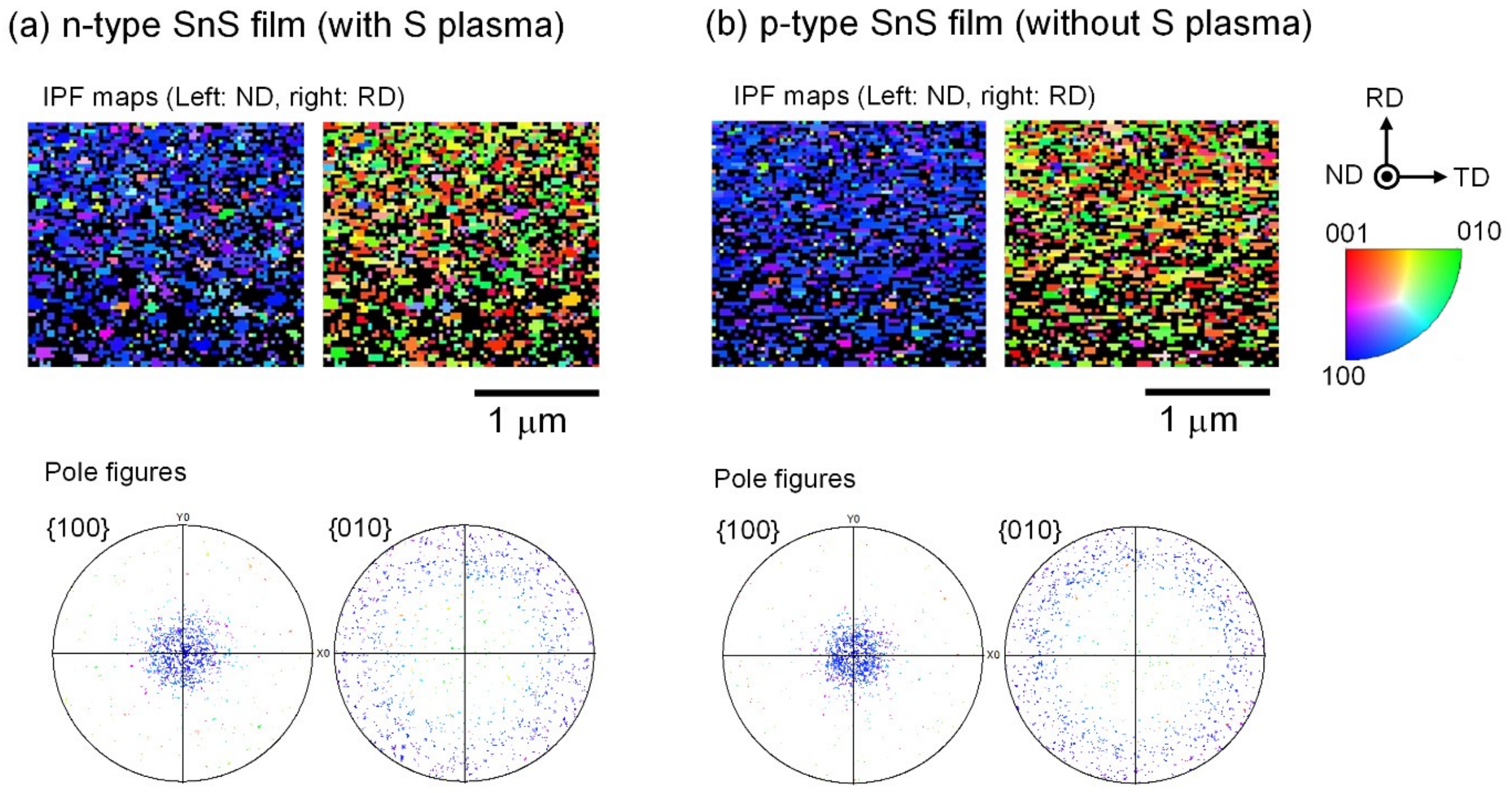

**Figure S4.** Top: IPF maps for ND and RD for (a) n-type and (b) p-type SnS thin films. Bottom: Pole figures of {100} and {010} for the areas in the maps. Colors in the pole figures correspond to those in the IPF maps for the ND.

## Section S5. In-plane analysis by TOF-SIMS measurement

**Figure S5** shows the in-plane distribution of secondary ions measured at 70–100 nm below the SnS thin film surface. All elements are uniformly distributed, and there is no segregation of elements at any specific position.

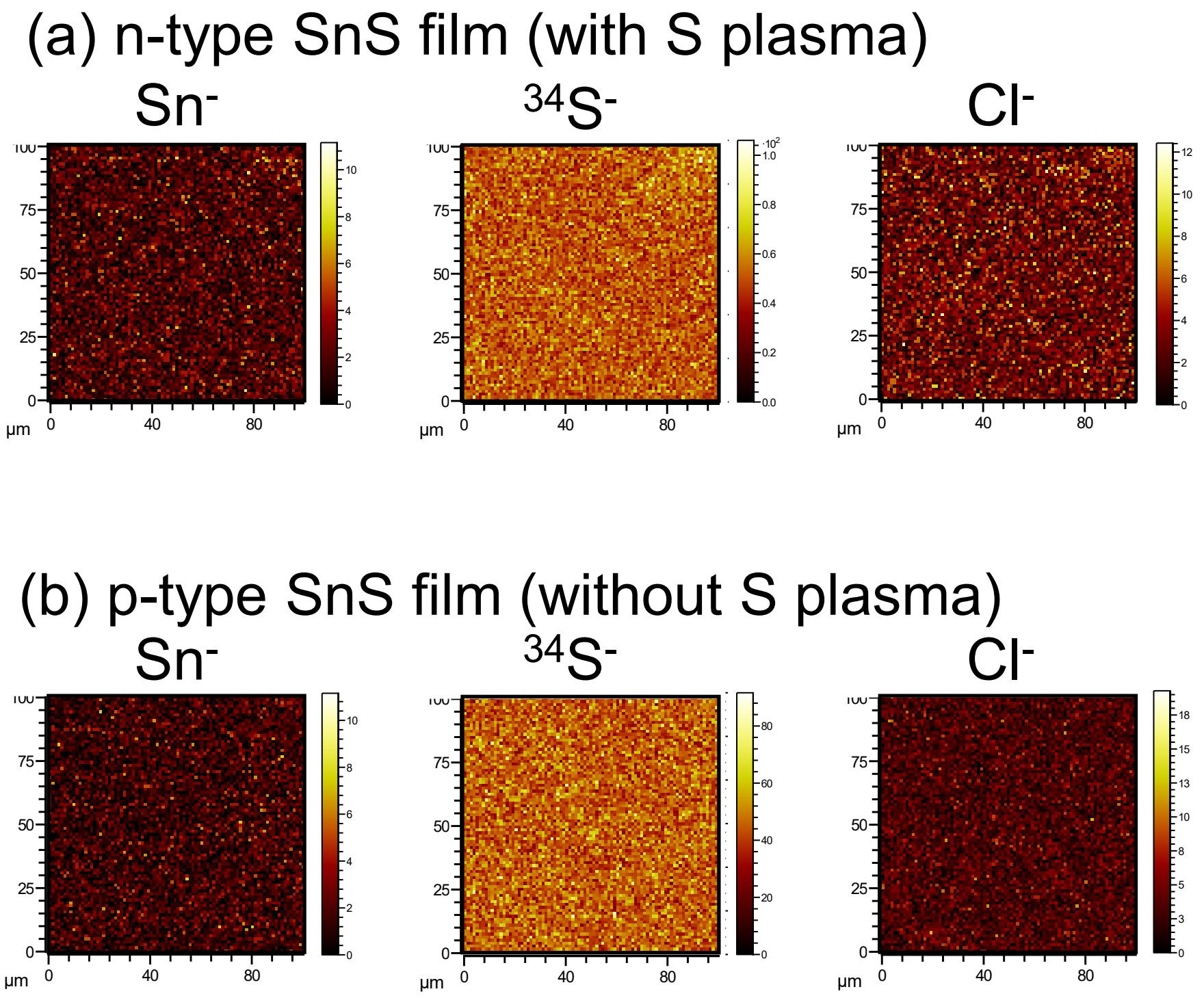

**Figure S5.** Secondary ion mapping of (a) n-type and (b) p-type SnS thin films deposited at $T_{sub}$ = 333 °C obtained by TOF-SIMS measurement.

## Section S6. Additional SEM and AFM images of thin film morphology

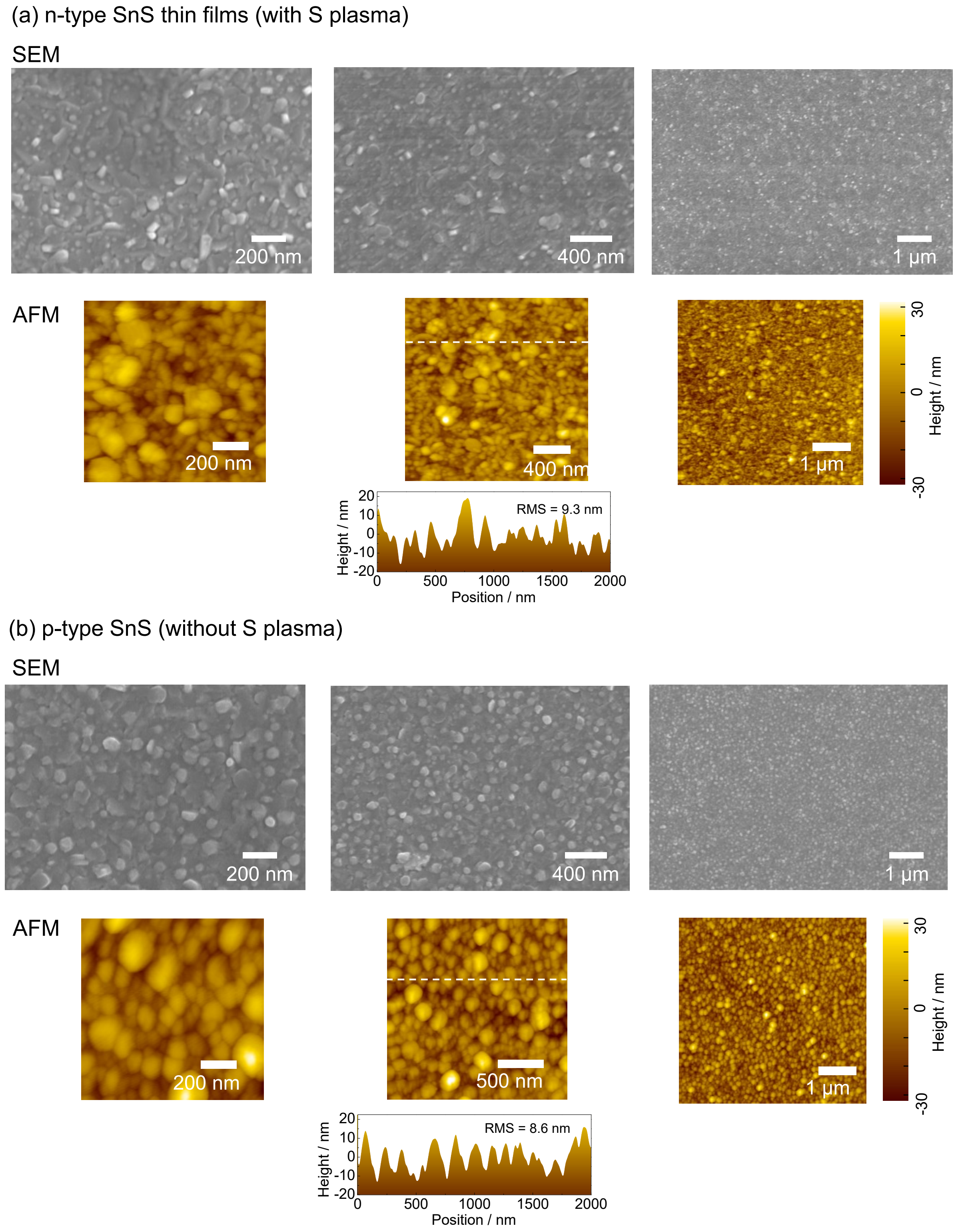

**Figure S6.** Surface morphological images observed by SEM and AFM for the films deposited at $T_{sub}$ = 333 °C (a) with and (b) without sulfur plasma supply. The AFM line profiles corresponding to the broken white lines in the images and the RMS roughness are also shown.

## Section S7. XPS with Ag-Lα excitation

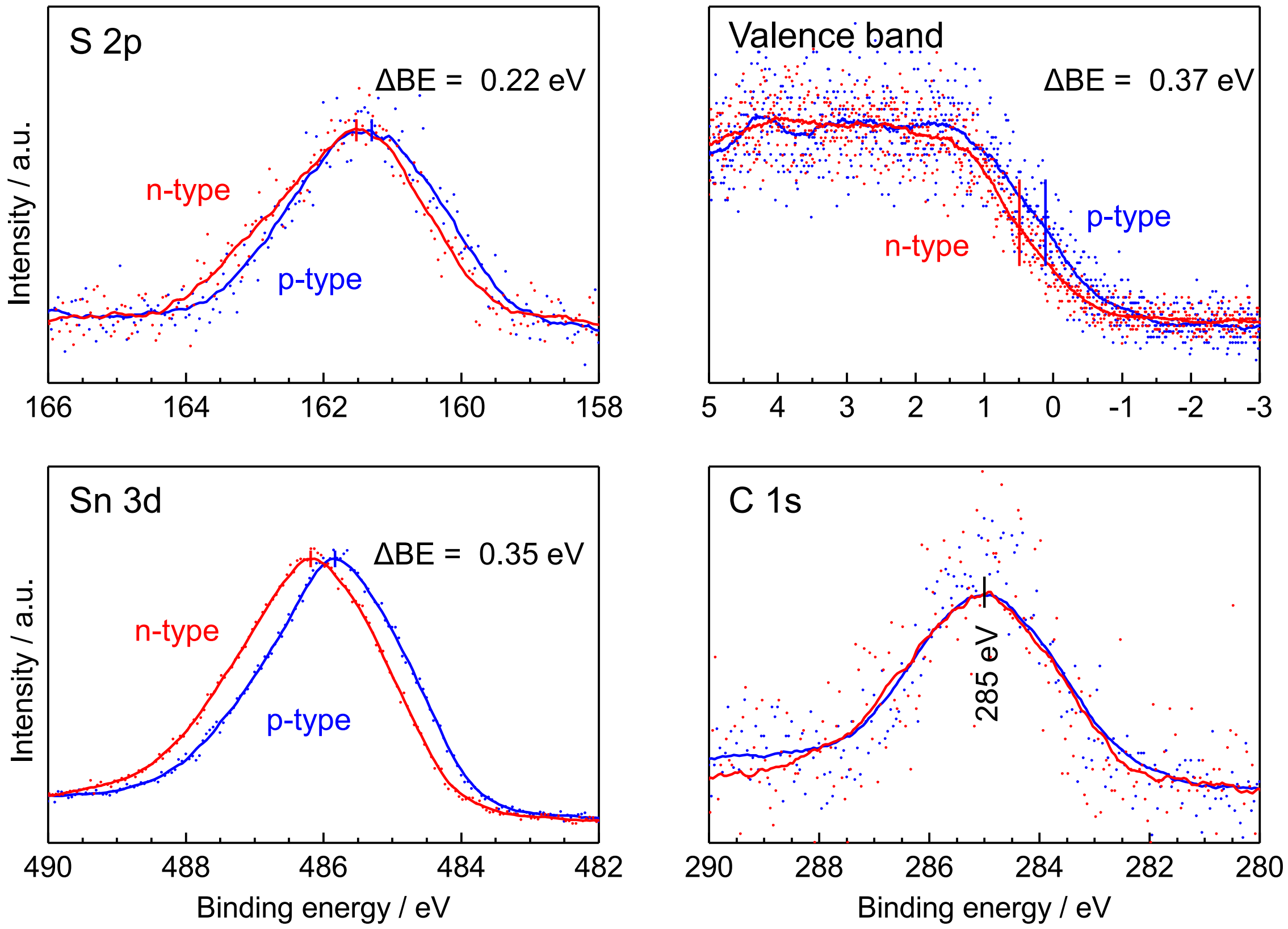

**Figure S7.** XPS spectra of the n-type and p-type films prepared at $T_{sub} = 222$ °C. Scattered data: measurement points, curves: moving average.

**Figure S7** shows the core level XPS spectra of the n-type and p-type SnS thin films prepared at a substrate temperature of 222 °C. The positions of S 2p and Sn 3d peaks of n-type films are ~300 meV higher than those of p-type films. Since the energies between the valence band maximum and the core levels are material constant[S1], this result indicates that the Fermi level of the n-type film is closer to the conduction band minimum as compared to the p-type film. Therefore, the carrier type of the obtained films was again confirmed in terms of their electronic structures. Note that while the valence band maximum is generally extracted by linear extrapolation of the leading valence band edge, it cannot be applied to this spectrum because of the poor total measurement resolution (~0.5 eV) that causes significant spectral broadening. Since the measurement was not in-situ, that is, the samples were exposed to the air before measurement, the spectra were affected to a certain extent by the top contamination and oxidation layers.

## Section S8. Optical properties and determination of band gaps

**Figure S8(a,b)** show the transmittance and absorption spectra of the obtained films. In **Figure S8(c,d)**, the indirect and direct band gaps were determined by the $h\nu$ vs. $(\alpha h\nu)^{1/r}$ plots, where $h\nu$, $\alpha$, and $r$ are the photon energy, absorption coefficient, and an index related to the inter-band transition type (2 for indirect allowed transition and 1/2 for direct allowed transition), respectively[S2] [28]. The transmittance at 1500 nm (0.83 eV) was corrected to 100% for Figure S9(c,d). The obtained band gaps are summarized in **Table S3**.

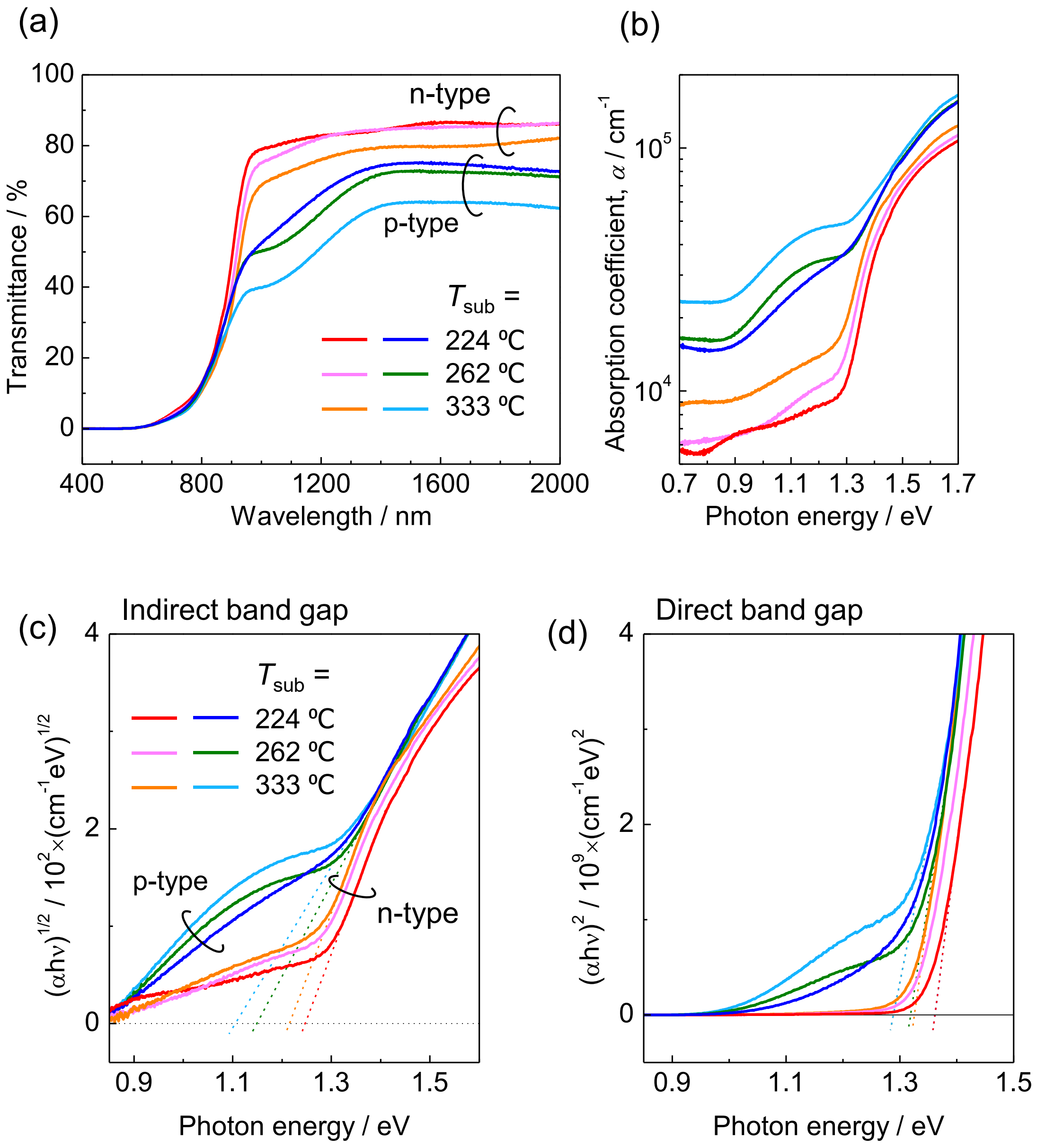

**Figure S8**. (a) Transmittance and (b) absorption spectra of films deposited at various substrate temperatures. (c) Indirect and (d) direct band gaps of the n-type and p-type SnS thin films.

**Table S3**. Indirect and direct band gaps (eV) of n-type and p-type films prepared at different substrate temperatures, based on data in Figure S8(c,d).

| $T_{sub}$ / °C | n-type (with sulfur plasma) | | p-type (without sulfur plasma) | |
|---|---|---|---|---|
| | Indirect gap | Direct gap | Indirect gap | Direct gap |
| 224 | 1.24 | 1.36 | 1.13 | 1.31 |
| 262 | 1.22 | 1.34 | 1.14 | 1.32 |
| 333 | 1.21 | 1.33 | 1.12 | 1.29 |

## Section S9. Double log plots of Hall mobility vs. temperature

**Figure S9** shows double log plots of Hall mobility vs. temperature of the SnS thin films prepared at a substrate temperature of 333 °C. The obtained slopes are +2.5 and +2.6 for the n-type and p-type films, respectively. These values are not consistent with cases dominated by ionized impurity scattering (+1.5) or phonon scattering (–1.5).

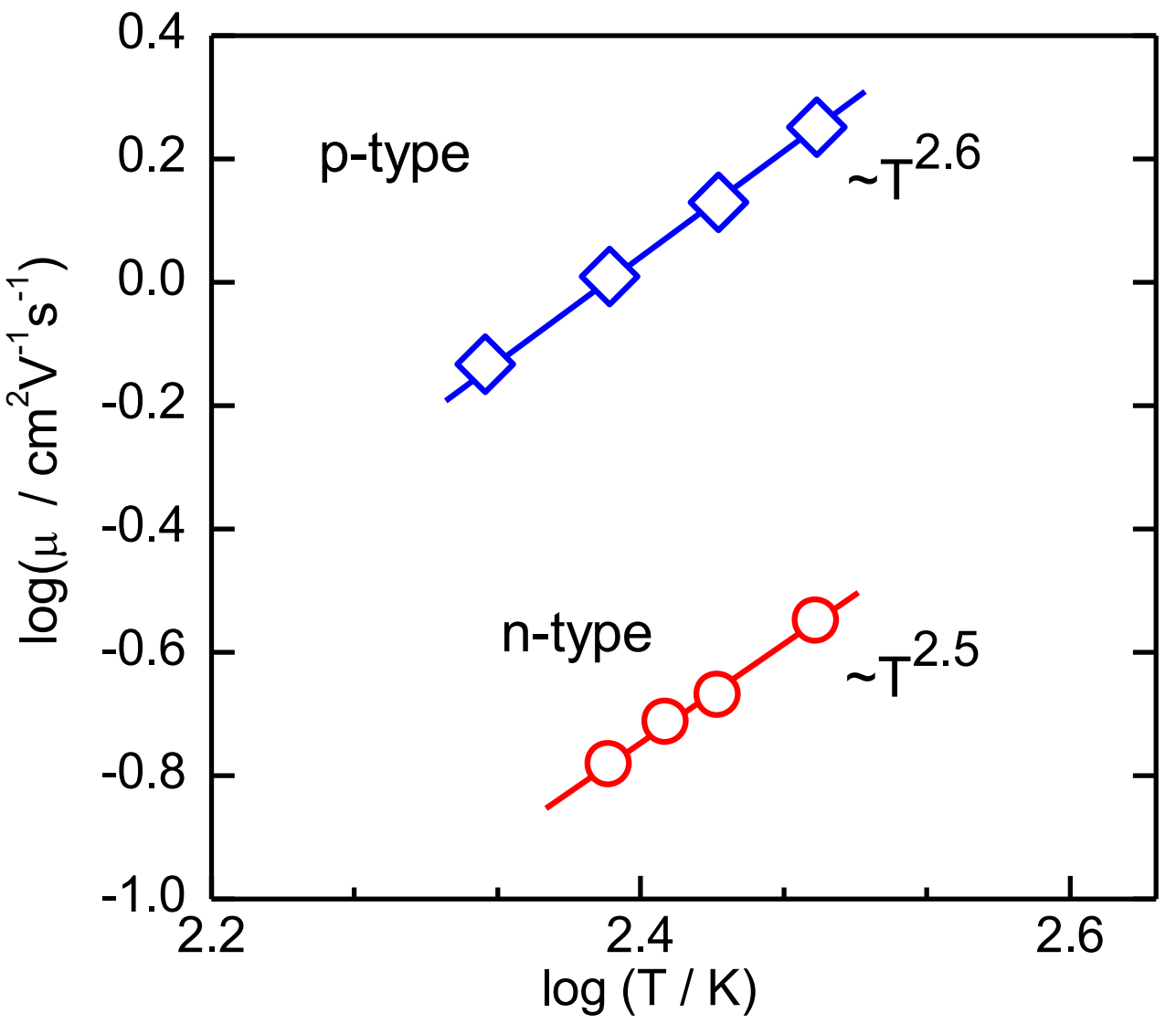

**Figure S9.** Double-log plots of Hall mobility and temperature of the SnS thin films prepared at $T_{sub}$ = 333 °C.

## Section S10. Detailed dependence of electrical properties on substrate temperature

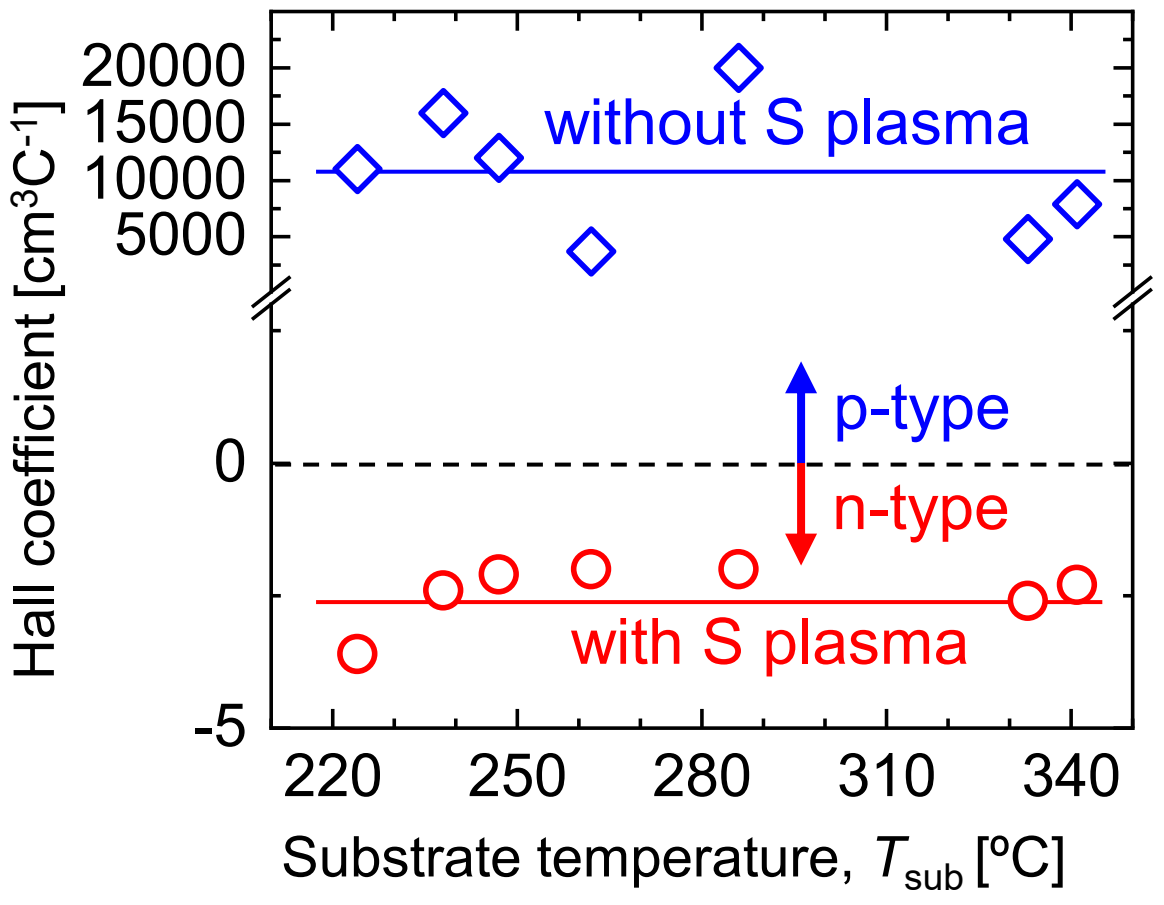

**Figure S10.** Substrate temperature dependence of the Hall coefficient of the SnS thin films deposited with and without sulfur plasma.

**Table S4.** Electrical properties of the n-type films (i.e., prepared with sulfur plasma) deposited at different substrate temperatures.

| $T_{sub}$ / °C | Hall coeff. /cm$^3$ C$^{-1}$ | Carrier type | $N$ / cm$^{-3}$ | $\mu$ / cm$^2$ V$^{-1}$ s$^{-1}$ | $\sigma$ / S cm$^{-1}$ |
|---|---|---|---|---|---|
| 341 | $-2.3\times10^{0}$ | n | $2.7\times10^{18}$ | $9.8\times10^{-2}$ | $4.3\times10^{-2}$ |
| 333 | $-2.6\times10^{0}$ | n | $2.4\times10^{18}$ | $2.4\times10^{-1}$ | $9.2\times10^{-2}$ |
| 286 | $-2.0\times10^{0}$ | n | $3.1\times10^{18}$ | $8.2\times10^{-1}$ | $4.1\times10^{-1}$ |
| 262 | $-2.0\times10^{0}$ | n | $3.1\times10^{18}$ | $1.1\times10^{0}$ | $5.6\times10^{-1}$ |
| 247 | $-2.1\times10^{0}$ | n | $2.9\times10^{18}$ | $5.3\times10^{-1}$ | $2.5\times10^{-1}$ |
| 238 | $-2.4\times10^{0}$ | n | $2.6\times10^{18}$ | $2.6\times10^{-1}$ | $1.1\times10^{-1}$ |
| 224 | $-3.6\times10^{0}$ | n | $1.8\times10^{18}$ | $1.3\times10^{-1}$ | $3.6\times10^{-2}$ |

**Table S5.** Electrical properties of the p-type films (i.e., prepared without sulfur plasma) deposited at different substrate temperatures.

| $T_{sub}$ / °C | Hall coeff. / cm$^3$ C$^{-1}$ | Carrier type | $N$ / cm$^{-3}$ | $\mu$ / cm$^2$ V$^{-1}$ s$^{-1}$ | $\sigma$ / S cm$^{-1}$ |
|---|---|---|---|---|---|
| 341 | $7.9\times10^{3}$ | p | $8.0\times10^{14}$ | $1.9\times10^{0}$ | $2.4\times10^{-4}$ |
| 333 | $4.8\times10^{3}$ | p | $1.4\times10^{15}$ | $1.5\times10^{0}$ | $3.1\times10^{-4}$ |
| 286 | $2.0\times10^{4}$ | p | $3.2\times10^{14}$ | $1.8\times10^{0}$ | $9.3\times10^{-5}$ |
| 262 | $3.7\times10^{3}$ | p | $1.7\times10^{15}$ | $7.0\times10^{-1}$ | $1.9\times10^{-4}$ |
| 247 | $1.2\times10^{4}$ | p | $5.1\times10^{14}$ | $1.0\times10^{0}$ | $8.5\times10^{-5}$ |
| 238 | $1.6\times10^{4}$ | p | $3.9\times10^{14}$ | $1.1\times10^{0}$ | $6.6\times10^{-5}$ |
| 224 | $1.1\times10^{4}$ | p | $5.7\times10^{14}$ | $8.6\times10^{-1}$ | $7.9\times10^{-5}$ |

## Section S11. XRD profiles of SnS thin films deposited at various substrate temperatures

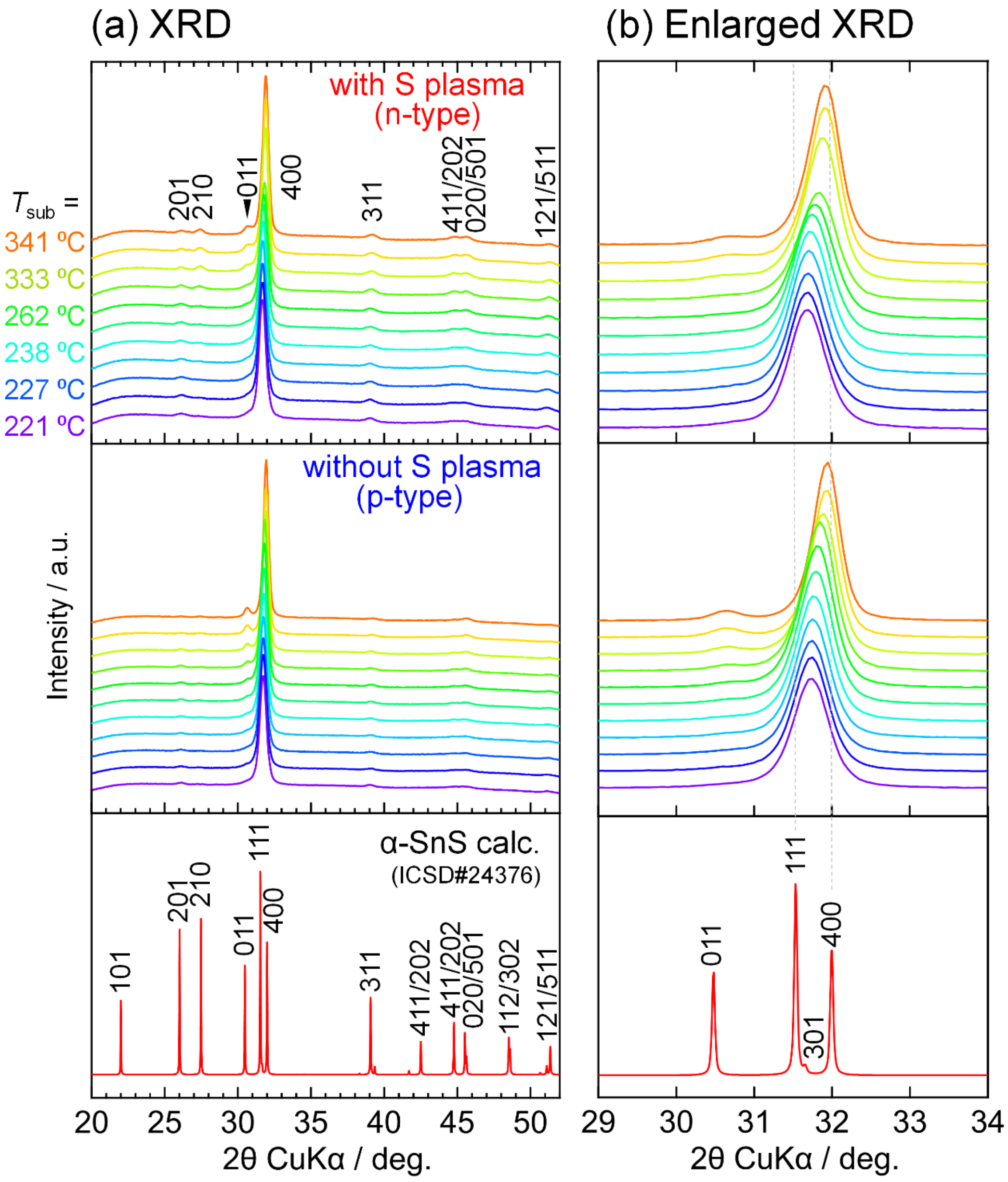

**Figure S11.** XRD characterization of phases in the films deposited at various substrate temperatures, together with the reference pattern of α-SnS (ICSD# 24376) in the range of (a) 2θ = 20–52° and (b) 29–34°. The corresponding $T_{sub}$ are shown on the left side.

## Section S12. In-plane XRD measurements of the thin films deposited at various substrate temperatures

In-plane analysis of the obtained SnS thin films were performed with X-ray diffractometer (XRD, Smartlab, Rigaku, Japan). The incident angle was 0.4 –0.5°.

The EBSD measurement (Figure S4) indicated that the films prepared at high temperature ($T_{sub}$ = 333 °C) possessed the preferential (100)-orientation. Therefore, the primary peaks with dominant intensity in the out-of-plane profiles of the films prepared at the high temperature are 400-diffractions (Figure S12(a,c)). On the other hand, in their in-plane profiles, 400-diffractions were no longer dominant, and 011-diffraction became significantly stronger. Since 011-plane is perpendicular to 400-plane, these results are well consistent with the fact that they are (100)-oriented films. Similar characteristics were observed for the films prepared at $T_{sub}$ = 224 °C (Figure S12(b,d); the primary peak with dominant intensity was observed in the out-of-plane profiles, which was then no longer dominant in the in-plane profiles and 011-diffraction became significant at the same time. Therefore, the primary peaksof the films prepared at $T_{sub}$ = 224 °C in the out-of-plane profiles should also be 400-diffractions, and its shift depending on the substrate temperature (Figure S11(b)) indicates the change of the lattice constant of the *a*-axis ($a_0$). The substrate temperature dependence of $a_0$ is shown as Figure 4(a) in the main text.

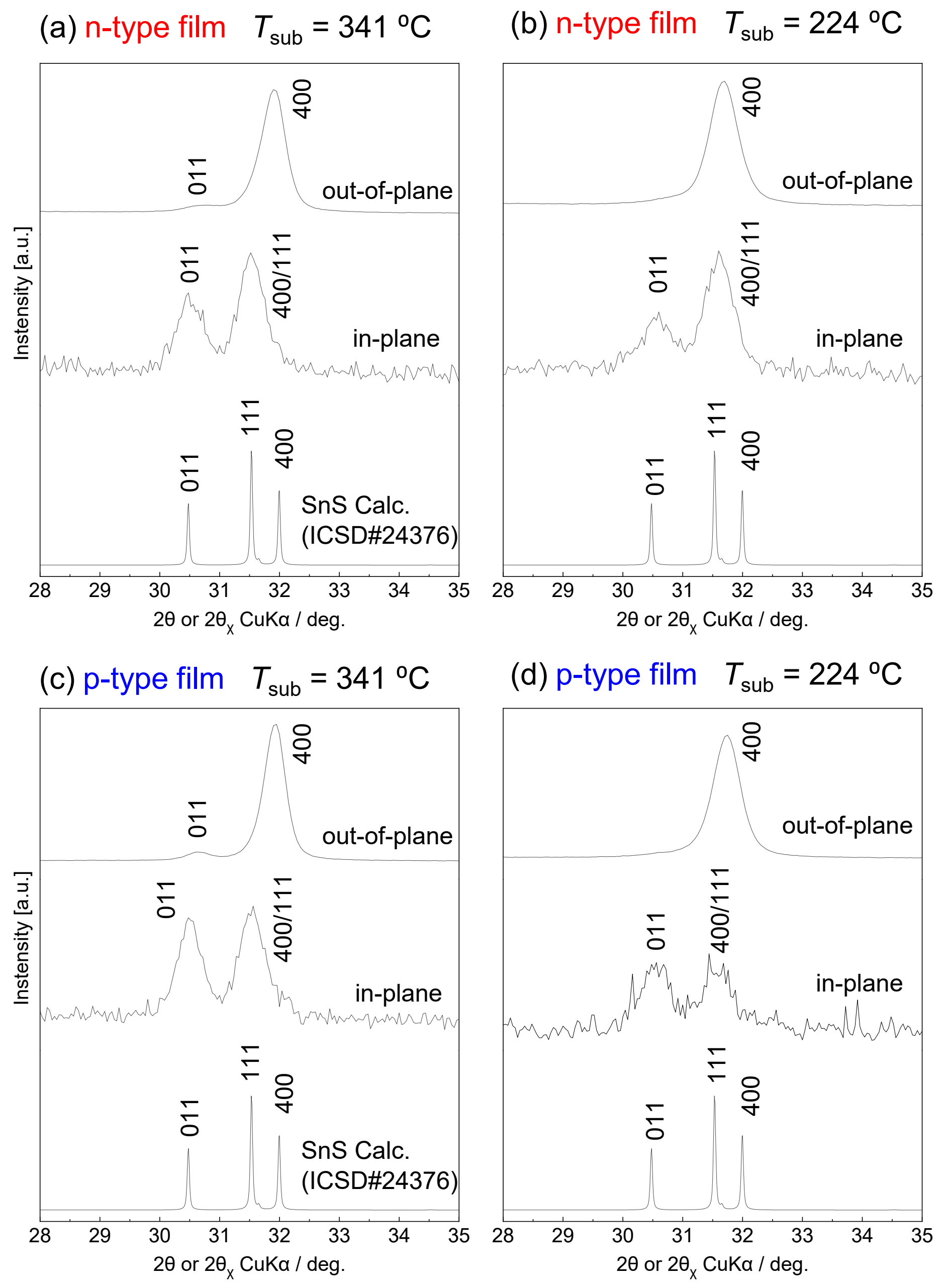

Figure S12. The out-of-plane and in-plane profiles of the SnS thin films prepared different substrate temperature together with the calculated SnS powder patterns (ICSD#24376). The out-of-plane ones are enlarged reprint of Figure S11.

## Section S13. Thickness of the films

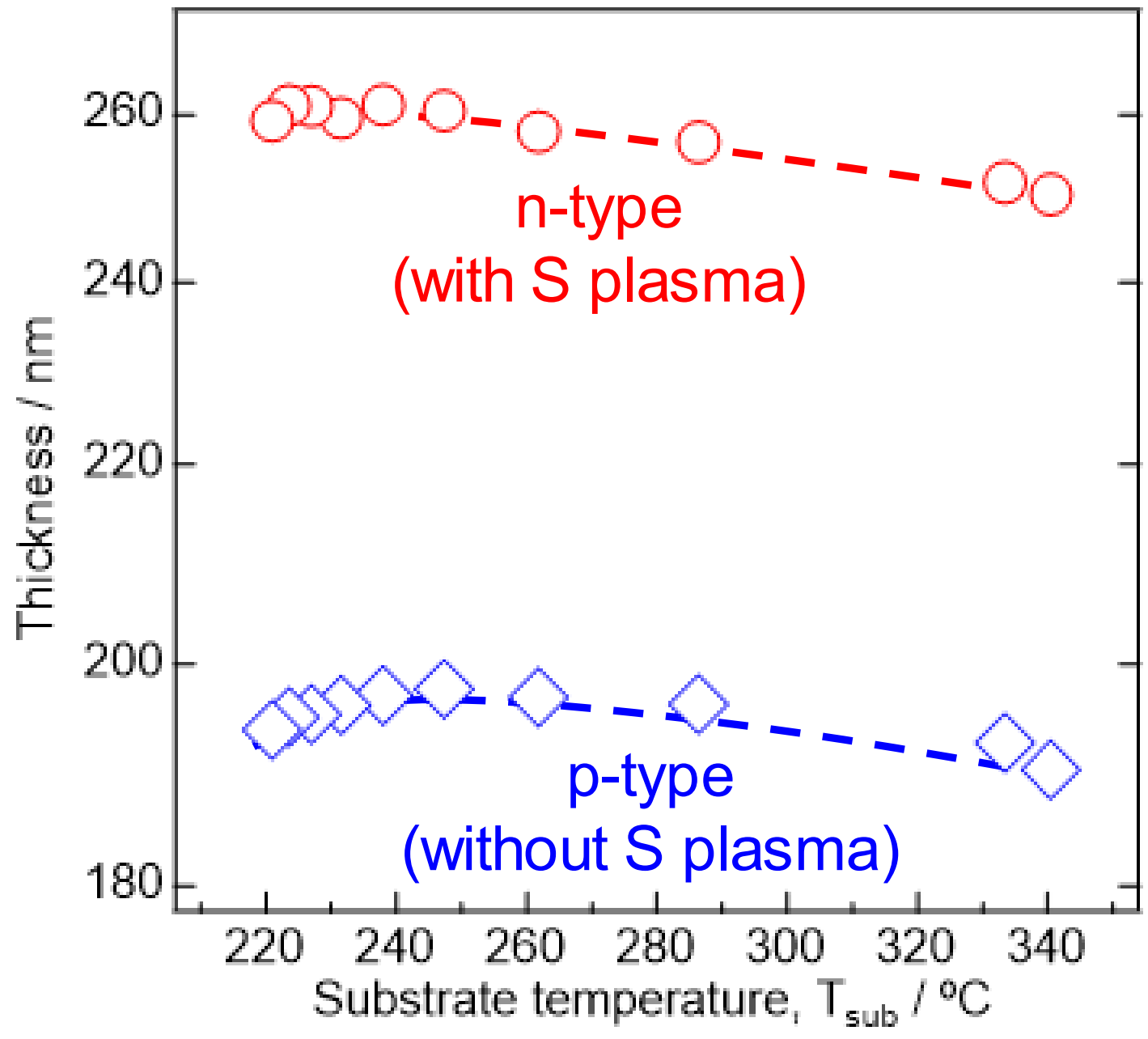

**Figure S13.** Film thickness as a function of substrate temperature by XRF measurements.

## References for supporting information.